\documentstyle[prl,aps]{revtex}

\input{psfig.sty}

\begin{document}
\draft

\twocolumn[
\hsize\textwidth\columnwidth\hsize\csname @twocolumnfalse\endcsname
\title{Topological defects and Goldstone excitations in domain walls between ferromagnetic
quantum Hall effect liquids}
\author{Vladimir I. Fal'ko $^{\ast \dagger }$, S.V. Iordanskii $^{\ddagger \ast }$}
\address{$^{\ast }$ School of Physics and Chemistry, Lancaster University, LA1
4YB, UK\\
$^{\dagger }$ Departement de Physique, Universit\'{e} Joseph Fourier Grenoble
I, Grenoble, France \\
$^{\ddagger }$ Landau Institute for Theoretical Physics, ul. Kosygina 1,
Moscow, Russia}
\date{today}

\maketitle

\begin{abstract}
It is shown that the low-energy spectrum of a ferromagnetic quantum Hall
effect liquid in a system with a multi-domain structure generated by an
inhomogeneous bare Zeeman splitting $\varepsilon _{Z}$ is formed 
by excitations localized at the walls
between domains. For a step-like $\varepsilon _{Z}({\bf r})$, the domain wall 
spectrum includes a spin-wave with a linear dispersion and a small 
gap due to spin-orbit coupling, and a low-energy topological
defects.  The latter are charged and may dominate in the transport under 
conditions that the
percolation through the network of domain walls is provided.
\end{abstract}
\pacs{73.40.Hm, 64.60.Cn, 75.10.-b}
\narrowtext] 
Due to the electron-electron exchange interaction,
two-dimensional (2D) electrons form a ferromagnetic liquid \cite{Girvin} in
the vicinity of odd-integer filling factors. Recently, an idea has been
proposed \cite{Lee,Sondhi,MacDonald} that the dissipation and thermodynamics
of the liquid at the filling factor $\nu =1$ may be dominated by a
topological spin-texture (skyrmion) which provides a non-trivial degree of
mapping of a 2D plane onto a sphere representing the local direction of
spin-polarization $\overrightarrow{n}$, $n=1$. As a particle-like
excitation, the skyrmion carries electrical charge equal to its topological
charge, and has a reduced contribution of the exchange to its energy \cite
{Sondhi}, as compared to the dissociation energy of a spin-exciton into an
electron-hole pair \cite{BychIordElia}. It has been suggested \cite
{Sondhi,Eisenstein} that the conductivity $\sigma _{xx}$ at exactly $\nu =1$
may be provided by thermal activation of skyrmion/anti-skyrmion pairs.
However, the excess Zeeman energy of a skyrmion may compensate the gain in
the exchange energy, due to a large number of improperly polarized electrons
in it. To stabilize skyrmions, one should reduce the bare value of the
single-electron Zeeman splitting, $\varepsilon _{Z}$ in the material
accommodating the 2D electron gas. In GaAs/AlGaAS heterostructures and
quantum wells, one may achieve this goal either by using the hydrostatic
pressure \cite{Maude,Leadley} which induces strains and progressively
changes the conduction band Land\'{e} factor from negative to positive
values, or by optically polarizing nuclear spins \cite{Kukunmr} of Ga and As
in the vicinity of the heterostructure, which affects $\varepsilon _{Z}$ via
the hyperfine interaction \cite{Dobers}. Both of these two methods have
recently allowed one to obtain the 2D electron gas with an almost zero value
of the single-electron Zeeman splitting \cite{Maude,Leadley,Kukunmr},
although it is not clear to which extend the induced variation of $%
\varepsilon _{Z}$ is spatially homogeneous over the plane of the 2D gas.

In this paper, we study the effect of an inhomogeneity of the
single-electron spin splitting onto the ferromagnetic quantum Hall effect
liquid at $\nu =1$ under the extreme condition that the externally modified
Zeeman energy approaches zero value and extend an analysis of
skyrmionic-type textures onto the latter case. For the sake of simplicity,
we consider a strained quantum well \cite{Maude} in which $\left\langle
\varepsilon _{Z}({\bf r})\right\rangle =0$, but only as an average over the
2D plane, whereas locally it fluctuates between small positive and negative
values, $\left\langle \varepsilon _{Z}^{2}({\bf r})\right\rangle
=\varepsilon _{Z}^{2}$. Mesoscopic-size regions of a negative and positive $%
\varepsilon _{Z}$ generate a multi-domain structure of the ferromagnetic
liquid. A soft disorder in the sign of the Zeeman energy is different from
the traditionally studied potential disorder, since the former cannot be
electrostatically screened. In a given sample, the lines where $\varepsilon
_{Z}({\bf r})=0$ define a contour of walls between oppositely polarized
domains - their spatial configuration is determined by a distribution of
stresses and by the interface roughness in a given quantum well device \cite
{Maude,Leadley}, or by inhomogeneity of the nuclear spin polarization \cite
{Kukunmr}. Below, we analyze the spin-texture of a domain wall between two
oppositely polarized $\nu =1$ ferromagnetic liquids (DW), its excitations
and a possible effect of topological defects in the DW on the transport
properties of the 2D system.

Several features of a domain wall between quantum Hall effect ferromagnets
can be assessed qualitatively. Due to the stiffness of a polarization field,
it can be regarded as a smooth variation of $\overrightarrow{n}({\bf r})$
created by conditions imposed by the crystal: from $n_{z}(x>0)=1$ to $%
n_{z}(x<0)=-1$. The long-range character of the variation of the order
parameter $\overrightarrow{n}({\bf r})$ across the DW, at the length scale $%
L_{w}$ $\gg \lambda =\sqrt{\hbar c/eB}$, is guaranteed by the smallness of
the Zeeman energy as compared to the electron-electron exchange, $%
\varepsilon _{Z}\ll \Im _{0}$. A spin-texture inside the domain wall can be
described as a mapping of the 2d plane on to the unit sphere, such that $%
\overrightarrow{n}({\bf r})$ retraces a path from the 'south' to the 'north'
pole when a path in the real space traverses the border between domains. In
the case of a flat border between $\varepsilon _{Z}>0$ and $\varepsilon
_{Z}<0$ regions, the geodesics on a sphere naturally minimize the DW
exchange energy.

Without spin-orbit coupling, the DW energy would be degenerate with respect
to the choice of a geodesic, i.e., with respect to the angle $\varphi $
between the planar component of the polarization vector $\overrightarrow{n}%
=(\sin \nu \cos \varphi ,\sin \nu \sin \varphi ,\cos \nu )$ and the locally
determined normal direction to the line\ where $n_{z}=\cos \nu =0$. This
degeneracy results in a soft collective mode corresponding to the rotation
of the spin coordinate system relative to the orbital one. \ Symmetry
transformations responsible for this residual degeneracy belong to the group 
$SO_{2}$ of rotations of a spin-texture around the axis $\overrightarrow{l}%
_{z}$ - the magnetic field orientation \cite{Cote,Karlhede}. The spin-orbit
coupling of a form proposed in \cite{BychRashba}\ lifts the degeneracy of
the DW ground state and orients the field $\overrightarrow{n}$ in the middle
of the DW perpendicular to the border between domains. As a one-dimensional
object, the DW has an action that resembles the action of a classical
sine-Gordon model \cite{Fradkin,TsvelikBook}. Its low-lying excited states
have the form of (a) spin-waves and (b) of topologically stable defects -
kinks in the angle $\varphi $.

The ground state of a single straight domain wall is characterized by one
properly chosen geodesic line on a sphere $\left| \overrightarrow{n}\right|
=1$. Spin-waves correspond to small rotations, $\varphi \ll 1$, of a
geodesic. A kink can be described as a $\Delta \varphi =2\pi $ rotation of
the polarization vector $\overrightarrow{n}$ around the axis $%
\overrightarrow{l}_{z}$. Topological classes of kinks are given by the
homotopic group $\pi _{1}(SO_{2})=Z$, or, equivalently, by the degree of
mapping of a line representing the middle of the DW, $n_{z}=0$ onto the
equator of a unit sphere. As a 2D object, a kink can be viewed as a mapping
of a plane into a sphere, $R^{2}\rightarrow S^{2}$, such that a set of
geodesics collected upon moving along the domain wall covers the entire
sphere $\left| \overrightarrow{n}\right| =1$. Since the excess density of
electrons, $\delta \rho $ carried by a polarization texture is equal to \cite
{Lee,Sondhi} 
\begin{equation}
\delta \rho =\left[ \left( \partial _{y}\cos \nu \right) \left( \partial
_{x}\varphi \right) -\left( \partial _{x}\cos \nu \right) \left( \partial
_{y}\varphi \right) \right] /4\pi ,  \label{eq1}
\end{equation}
the electrical charge of a kink is equal to the degree of mapping $%
R^{2}\rightarrow S^{2}$ provided by the defect, which coincides with its
topological charge classified by $\pi _{1}(SO_{2})$. That makes kinks akin
skyrmions, though the latter are classified using the homotopic group $\pi
_{2}(SU_{2}/U_{1})=Z$, and these two solitonic excitations are related to
different boundary conditions at the infinities of a 2D plane. Stability of
a kink with respect to the decay of its charge density into the bulk is
provided by its relatively small energy, as compared to the bulk skyrmion,
which is supported by the \ quantitative analysis. The kinks we discuss in
this paper are also similar to skyrmionic textures on the edge of the
quantum Hall effect liquid discussed in \cite{Karlhede}.

The quantitative analysis of the DW energetics in this paper is based on the 
$\sigma $-model approach \cite{Fradkin}. We started from the Grassman
functional integral for interacting electrons where the kinetic energy,
Zeeman splitting and spin-orbit coupling terms were present. Then, we split
interactions by means of the Hubbard-Stratonovich transformation, and
derived the 2D $\sigma $-model for the polarization $\overrightarrow{n}%
=(\sin \nu \cos \varphi ,\sin \nu \sin \varphi ,\cos \nu )$ using the
gradient expansion \cite{FIunpub}. Below, we parametrize the polarization by
using Euler angles: $\nu ,$ $|\nu |<\pi $, with respect to the normal to the
plane and $\varphi $ with respect to the direction across the domain wall.
The mean field $\overrightarrow{n}({\bf r},t)$ enters into a unitary
transformation, $U=\left[ \left( 1+n_{z}\right) ^{1/2}+i\left(
1+n_{z}\right) ^{-1/2}(n_{y}\sigma _{x}-n_{x}\sigma _{y})\right] /\sqrt{2}$,
which locally determines the spinor part of the electron wave function. The
gradient expansion with respect to the matrices $U(-i\lambda \nabla
)U^{\dagger }$, $U(-i\partial _{t})U^{\dagger }$ was based on the assumption
that the chemical potential in the system lies in the gap, and we took into
account the Landau level mixing both by interactions and the spin-orbit
coupling. That brings us to the action in the form 
\begin{equation}
\int \left[ F_{t}+(F_{\nabla }+F_{{\rm C}})+F_{Z}+F_{{\rm so}}\right] d{\bf %
r/}\left( 2\pi \right) -{\cal F}_{{\rm top}},  \label{lagran}
\end{equation}
which is composed of the gradient terms $(F_{\nabla }+F_{{\rm C}})$, the
Zeeman energy $F_{Z}$, the spin-orbit coupling term $F_{{\rm so}}$, the
topological term ${\cal F}_{{\rm top}}=(\mu -\Im _{1}/2)\int d{\bf r}\delta
\rho $, where $\mu $ is the single-electron chemical potential, and $%
F_{t}=-(i/2\lambda ^{2})\cos \nu \,\partial _{t}\varphi $. The latter term
should be kept to calculate the spin-wave spectrum of the DW \cite
{RemDtsquare}. The polarization field stiffness and the Zeeman energy are
described by

\begin{eqnarray}
F_{\nabla } &=&\Im _{1}\sum_{i}\left( \nabla n_{i}\right) ^{2}/8=\Im
_{1}\left( \sin ^{2}\nu \left( \nabla \varphi \right) ^{2}+\left( \nabla \nu
\right) ^{2}\right) /8\;  \nonumber \\
F_{Z} &=&\left( \varepsilon _{Z}/2\lambda ^{2}\right) \,{\rm sign}%
(x)n_{z}=\left( \varepsilon _{Z}/2\lambda ^{2}\right) \,{\rm sign}(x)\cos
\nu .  \label{eq3}
\end{eqnarray}
The Zeeman energy in Eq. (\ref{eq3}) corresponds to the model of a step-like 
$\varepsilon _{Z}({\bf r})=\varepsilon _{Z}{\rm sign}(x)$. The exchange
interaction of electrons at the n-th Landau level with electrons from a
completely filled Landau level n=0 is given by $\Im _{n}=\int_{0}^{\infty
}V(R)e^{-R^{2}/2}{\rm L}_{n}(R^{2}/2)RdR$, where $R=r/\lambda $. For an
unscreened Coulomb interaction, $V(r)=e^{2}/r\chi $, $\Im _{1}=\Im _{0}/2$,
and $\Im _{0}=\sqrt{\pi /2}e^{2}/\lambda \chi $. \ The 'Coulomb term' $F_{%
{\rm C}}$ in Eq. (\ref{lagran}) is the result of the higher order expansion
in gradients \cite{Sondhi}. It is mentioned only to be neglected later,
since $F_{\nabla }$ is enough to provide an ultraviolet cut-off in the DW
defect energy calculation.

In the main approximation, that is, before the spin-orbit coupling is taken
into account, the domain wall energy is minimized by any texture with $%
\varphi ={\rm const}$ and with $\nu (x)$ obeying the saddle-point equation, 
\begin{equation}
\lambda ^{2}\partial _{x}^{2}\nu =\left( 4\varepsilon _{Z}/\Im _{0}\right) 
{\rm sign}(x)\sin \nu .  \label{eqsaddlepoint}
\end{equation}
Eq. (\ref{eqsaddlepoint}) is the result of the variational principle applied
to the energy $F_{\nabla }+F_{Z}$. Its solution should be antisymmetric, so
that in the half-plane $x>0$ it satisfies the boundary conditions cos$\nu
(0)=0$ and cos$\nu (\infty )=1$. The optimal $\nu _{0}(x)$ can be found in
the form of 
\begin{equation}
\cos \nu _{0}=\left[ 1-2{\rm \cosh }^{-2}\left( |w|+\ln (\sqrt{2}+1)\right) %
\right] {\rm sign}(w),\;  \label{eqnu0}
\end{equation}
where $w=(2x/\lambda )\sqrt{\varepsilon _{Z}/\Im _{0}}$, which confirms that
the domain wall width, $L_{w}=2\lambda \sqrt{\Im _{0}/\varepsilon _{Z}}$, is
large, $L_{w}\gg \lambda $. The result of Eq.(\ref{eqnu0}) allows us to
calculate the ground state DW energy per magnetic length: $E_{w}^{0}=\sqrt{%
\varepsilon _{Z}\Im _{0}}(1-\sqrt{1/2})/\pi $ \cite{secondsaddle}.

At this scale of energies, the structure of the wall is degenerate with
respect to the choice of a geodesic on the unit sphere along which the
polarization $\overrightarrow{n}$ rotates, that is, with respect to the
angle $\varphi $. At a finer scale of energies, such a degeneracy is lifted
by the spin-orbit coupling. It is natural to assume that $\hbar \omega
_{c}\gg \Im _{0}=\frac{e^{2}}{\chi \lambda }\sqrt{\frac{\pi }{2}}\gg
\varepsilon _{Z}$, which confines the ground state electrons to the lowest
Landau level. Since the spin-orbit coupling \cite{BychRashba,Rossler}, such
as $v_{so}\left[ \overrightarrow{p}\times \overrightarrow{\sigma }\right] $,
is not diagonal in the Landau level basis, it appears only in the form of $%
\varepsilon _{so}\Im _{0}/\hbar \omega _{c}$, $\varepsilon _{so}=\hbar
v_{so}/\lambda $, which approves its perturbative treatment. To be more
specific, we shall limit the spin-orbit coupling energy by the constraint $%
\left( \varepsilon _{so}/\hbar \omega _{c}\right) ^{2}<\varepsilon _{Z}/\Im
_{0}$, which enables us to exclude a spin-wave instability of the ground
state of a homogeneous liquid in the bulk \cite{FIunpub}. The spin-orbit
coupling term in Eq. (\ref{lagran}) can be found in the form of 
\begin{eqnarray}
F_{{\rm so}} &=&\left( \varepsilon _{so}\Im _{1}/\hbar \omega _{c}\lambda
\right) n_{z}({\bf r})\partial _{i}n_{i}({\bf r})  \label{eqSO} \\
&=&\frac{\varepsilon _{so}\Im _{1}}{2\hbar \omega _{c}\lambda }\left\{ \cos
\varphi \partial _{x}+\sin \varphi \partial _{y}\right\} \left( \nu -\frac{%
\sin 2\nu }{2}\right) .  \nonumber
\end{eqnarray}
which has a minimum at $\varphi =0$ for the ground state of the DW with $\nu
({\bf r})=\nu _{0}(x)$ and determines the energies of the low-lying
excitations of this object.

First of all, we analyze the energetics of a topological defect. A kink
corresponds to the texture with the planar component of the polarization
(followed along the middle of the DW) making a full circle when the
coordinate changes from $y=-\infty $ to $y=\infty $. If the DW texture
varies along the wall much slower than across it - an assumption which can
be verified after the solution is found, the problem may be formulated as a
one-dimensional one. The free energy of the wall (regarded as a 1D system)
related to the lifted residual $SO_{2}$ degeneracy can be obtained from Eqs.
(\ref{eq3},\ref{eqSO}) by integrating out the transverse form-factor of the
DW, $\cos \nu _{0}(x)$: 
\[
{\cal E}[\varphi (y)]=\Im _{0}\int \frac{dy}{8\lambda }\left[ \sqrt{\frac{%
\Im _{0}}{\varepsilon _{Z}}}\left( \alpha \lambda \partial _{y}\varphi
\right) ^{2}+\frac{\varepsilon _{so}}{\hbar \omega _{c}}(1-\cos \varphi )%
\right] 
\]
where $\alpha =\sqrt{(2-\sqrt{2}/2)/3\pi }$ $\approx 0.\,\allowbreak 37$.
The energy minimum of a kink is provided by the texture with 
\[
\left\{ 
\begin{array}{c}
\cos \varphi (y)=1-\frac{2}{{\rm \cosh }^{2}\left( u\right) } \\ 
\sin \varphi (y)=\frac{2\Theta \sinh (u)}{{\rm \cosh }^{2}\left( u\right) }
\end{array}
\right. ,\;u=\frac{y}{\alpha \lambda }\left( \frac{\varepsilon _{so}}{\hbar
\omega _{c}}\sqrt{\frac{\varepsilon _{Z}}{\Im _{0}}}\right) ^{1/2}. 
\]
$\Theta =\pm 1$ is the sign of a topological charge of the defect. The
charge density distribution in a single kink calculated using Eqs. (\ref{eq1}%
) is shown in Fig. 1. The activation energy of the kink/anti-kink pair (two
times the spin-deformation energy of one kink) is equal to 
\begin{equation}
{\cal E}_{a}=\alpha \Im _{0}\left( \Im _{0}/\varepsilon _{Z}\right)
^{1/4}\left( \varepsilon _{so}/\hbar \omega _{c}\right) ^{1/2}.  \label{eqEa}
\end{equation}
Since the pair of defects with an opposite sign has total topological number 
$N=0$, the term ${\cal F}_{{\rm top}}$\ plays no role in determining ${\cal E%
}_{a}$. Note that the activation energy in Eq. (\ref{eqEa}) is smaller than
the activation energy of a skyrmion in the 2D bulk, ${\cal E}_{a}\ll \Im
_{0} $ (since we assume that $\left( \varepsilon _{so}/\omega _{c}\right)
^{2}<\varepsilon _{Z}/\Im _{0}$). This prevents the kink from decaying into
the bulk excitations.

The spectrum of spin-waves propagating along the DW can be found by
expanding the Lagrangian in the vicinity of the homogeneous ground state, $%
\nu =\nu _{0}(w)$ and $\varphi =0$ - up to the second order in small
variations, $\varphi =e^{iqy/\lambda -i\omega t/\hbar }\varphi (w)$ and $\nu
=\nu _{0}(w)+e^{iqy/\lambda -i\omega t/\hbar }\delta \nu (w)$ (the
coordinate across the wall is normalized by its width $L_{w}$, $%
w=(2x/\lambda )\sqrt{\varepsilon _{Z}/\Im _{0}}$). To select excitations
localized near the boundary between domains, one should study only those
solutions for which $\delta \nu (\pm \infty )=0$. In contrast, the boundary
conditions for $\varphi $ are free, since the fluctuations of $\varphi $
have no sense in the regions where $\sin \nu =0$. Therefore, it is easier to
formulate the eigenvalue problem for a different function, $\phi (w)=\sin
\nu _{0}(w)\varphi (w)$, which has more suitable boundary conditions $\phi
(\pm \infty )=0$.

After applying the variational principle to the result of such an expansion,
and in the limit of $q\ll \sqrt{\varepsilon _{Z}/\Im _{0}}$ (the wavelength
along the wall is much longer than the DW width), we arrive at the
eigenvalue equations, 
\begin{eqnarray}
i\omega \phi &\approx &\varepsilon _{Z}\left[ -\partial _{w}^{2}+{\rm sign}%
(w)\cos \nu _{0}\right] \delta \nu ,  \label{eqeigenmodes1} \\
-i\omega \frac{\delta \nu }{\Im _{0}} &\approx &\frac{\varepsilon _{Z}}{\Im
_{0}}\Pi \phi +\left[ \frac{q^{2}}{4}-\frac{2\varepsilon _{so}}{\hbar \omega
_{c}}\sqrt{\frac{\varepsilon _{Z}}{\Im _{0}}}\left[ \partial _{w}\nu _{0}%
\right] \right] \phi .  \label{eqeigenmodes2}
\end{eqnarray}
The operator on the right hand side of equation Eq. (\ref{eqeigenmodes1})
has a spectrum of eigenvalues with gap $u_{0}\sim 1$. Eigenvalues of the
operator $\Pi $ in Eq. (\ref{eqeigenmodes2}),

\[
\Pi =-\partial _{w}^{2}+\left( \partial _{w}^{2}\sin \nu _{0}\right) /\sin
\nu _{0}, 
\]
start from zero - for the function $\phi _{0}={\rm Const}\times \sin \nu
_{0}(w)$ -\ and are of the order of unity for other 'excited states' $\phi
_{n}$ \cite{FIunpub}. The lowest eigenvalue corresponds to the excitation
with $\varphi $ constant across the wall, but varying along it. This is the
Goldstone mode. Other excited states are spin-waves in the bulk of the
domain affected by the presence of a wall. Their spectrum starts above a gap
of the order of Zeeman energy, $\varepsilon _{Z}$. If we treat Eqs. (\ref
{eqeigenmodes1},\ref{eqeigenmodes2}) perturbatively and approximate the
lowest eigenstate by $\phi _{0}$, taking into account the expression on its
left-hand side in the diagonal approximation, we find that the soft
excitation has the dispersion 
\begin{equation}
\hbar \omega =\left( \pi \varepsilon _{Z}\Im _{0}/A\right) ^{1/2}\sqrt{%
(\varepsilon _{so}/\hbar \omega _{c})\sqrt{\varepsilon _{Z}/\Im _{0}}+\left(
\alpha q\right) ^{2}},  \label{eqdispersion}
\end{equation}
where $A=\sum_{k}\left| \int dwf_{k}(w)\sin \nu _{0}(w)\right|
^{2}/u_{k}\sim 1$ is determined by the normalized eigenstates, $f_{k}(w)$
and eigenvalues, $u_{k}$ of an equation $\left[ -\partial _{w}^{2}+\left|
\cos \nu _{0}\right| -u_{k}\right] f_{k}(w)=0$. So, the spectrum of a soft
mode has the gap $\Im _{0}\left( \varepsilon _{so}/\hbar \omega _{c}\right)
^{1/2}\left( \varepsilon _{Z}/\Im _{0}\right) ^{3/4}$ and a linear
dispersion $\ q\sqrt{\varepsilon _{Z}\Im _{0}}$ at \ $\left( \varepsilon
_{so}/\hbar \omega _{c}\right) ^{1/2}\left( \varepsilon _{Z}/\Im _{0}\right)
^{1/4}<q<\sqrt{\varepsilon _{Z}/\Im _{0}}$.

Note that, as a one-dimensional object, DW between oppositely polarized QHE
liquids resembles a 1D antiferromagnet. This analogy can be traced
throughout the properties of kinks and linear-dispersion spin-waves, and
also can be extended onto identification of regimes when classically
obtained solutions are sound. Classical treatment of spin-waves is provided
by their stability against decay into kink-antikink pairs, since ${\cal E}%
_{a}\gg \hbar \omega $. The latter may be compared to the known condition
for the border between the classical solution for the 1D Sin-Gordon model
and the Fermi-gas of kinks \cite{Fradkin,Coleman}: $\beta ^{2}/8\pi =(\pi
\alpha ^{2}\sqrt{A})^{-1}(\varepsilon _{Z}/\Im _{0})<1$. It has been also
shown previously \cite{Rodriguez}\ that the topological defect energy in an
antiferromagnet is lowered by quantum corrections arising from scattering of
spin-waves on it. One should, therefore, analyze similar corrections to the
DW-kink energy, ${\cal E}_{a}$. However, our estimation \cite{FIunpub}
indicates that those corrections do not alterate the relation ${\cal E}%
_{a}\gg \hbar \omega $.

As we have shown above, the domain wall (DW) has an excitation spectrum with
energies lower than the energies of excitations in the bulk. Therefore, the
low-gap spin-waves in it may be responsible for the low-frequency absorption
of a system in the regime when the multi-domain structure is formed.
Moreover, the fact that topological defect in the DW is charged and has a
low activation energy (since they are formed on the top of a texture imposed
by the external conditions), may lead to the dominating role of
kink/anti-kink pairs in the dissipative conductivity, $\sigma _{xx}$ when
the network of DW's forms a percolation cluster through the entire 2D plane.
Since DW's separate regions of a positive and negative value of $\varepsilon
_{Z}$, the critical regime is realized when the sample areas under the $%
\varepsilon _{Z}>0$ and $\varepsilon _{Z}<0$ domains are equal \cite
{EfrosShklovskii}, that is, when $\left\langle \varepsilon _{Z}({\bf r}%
)\right\rangle =0$. This scenario suggests that in the proximity of external
conditions providing $\left\langle \varepsilon _{Z}\right\rangle \rightarrow
0$, dissipative transport at $\nu =1$, $\sigma _{xx}(T)$ may be determined
by percolation of thermally activated kinks through an infinite cluster of
DW's. That may be the reason why, in the high-pressure experiment \cite
{Maude,Leadley}, in the vicinity of the pressure value where the conduction
band $g$-factor in GaAs should nominally change its sign, the $\nu =1$
activation energy falls sharply below the value expected for the
skyrmion/anti-skyrmion pairs in a homogeneous ferromagnetic liquid.

Authors thank A.Tsvelik and B.Halperin for illuminating discussions,
D.Maude, R.Nicholas and M.Potemski - for discussions of experiments. This
work was supported by EPSRC and RFFI (grant 98-02-16245).

\begin{figure}[tbp]
\centerline{\psfig{figure=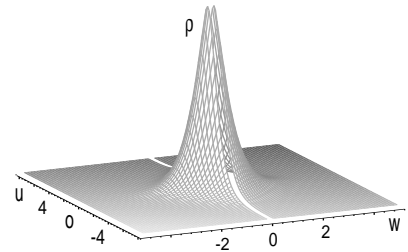,height=4cm,width=6cm}}
\caption{2D charge-density distribution around a kink. The scratch shows the
middle of the DW, where $g({\bf r})=0$. The coordinates across and along the
DW, $w$ and $u$, are normalized by $L_{w}$ and $L_{k}$ \protect\cite
{RemSingularity}, respectively.}
\end{figure}


\begin{references}
\bibitem{RemSingularity}  An isolated domain surrounded by a curved wall may
correspond to the same topological class as a skyrmion, so that it may be
charged, but we assume that the linear length scale, $L$ of the domain
structure is large enough to study the DW locally, as a 1D object. The
latter requires $L>L_{k}=\alpha \lambda \left( \hbar \omega _{c}/\varepsilon
_{so}\right) ^{1/2}\left( \Im _{0}/\varepsilon _{Z}\right) ^{1/4}$. The
relation $L_{k}>L_{w}$ is provided by the inequality $\left( \varepsilon
_{so}/\hbar \omega _{c}\right) ^{2}<\varepsilon _{Z}/\Im _{0}$.
\bibitem{RemDtsquare}  The higher-order terms in $\partial _{t}$ do not
affect dispersion.
\bibitem{secondsaddle}  The saddle point equation has also solution $\cos
\nu _{1}=\left[ 1-2{\rm \cosh }^{-2}\left( 2\left( x/\lambda \right) \sqrt{%
\varepsilon _{Z}/\Im _{0}}-\ln (\sqrt{2}+1)\right) \right] {\rm sign}(x),$
with $E_{w}^{1}=\sqrt{\varepsilon _{Z}\Im _{0}}(1+\sqrt{1/2})/\pi
>E_{w}^{0}. $
\end{references}
\end{document}